\documentclass[]{aastex631}

\begin{document}

\title{Multi-Mission Observations of Relativistic Electrons and High-Speed Jets Linked to Shock Generated Transients}


\correspondingauthor{Savvas Raptis}
\email{savvas.raptis@jhuapl.edu/savvasraptis@pm.me}

\author[0000-0002-4381-3197]{Savvas Raptis}
\affiliation{The Johns Hopkins Applied Physics Laboratory, Laurel, MD, USA}

\author[0000-0001-6342-225X]{Martin Lindberg}
\affiliation{Department of Physics and Astronomy, Queen Mary University of London, Mile End Road, London E1 4NS, UK}

\author[0000-0003-1778-4289]{Terry Z. Liu}
\affiliation{Department of Earth, Planetary, and Space Science, University of California, Los Angeles, USA}

\author[0000-0002-4381-3197]{Drew L.  Turner}
\affiliation{The Johns Hopkins Applied Physics Laboratory, Laurel, MD, USA}

\author[0000-0001-8532-3076]{Ahmad Lalti}
\affiliation{Northumbria University, Newcastle upon Tyne, UK}

\author[0000-0002-7964-7338]{Yufei Zhou}
\affiliation{School of Science, Harbin Institute of Technology (Shenzhen), Shenzhen, China}

\author[0000-0002-0625-8892]{Primo\v{z} Kajdi\v{c}}
\affiliation{Departamento de Ciencias Espaciales, Instituto de Geofísica, Universidad Nacional Autónoma de México, Mexico City, Mexico}

\author[0000-0001-6589-4509]{Athanasios Kouloumvakos}
\affiliation{The Johns Hopkins Applied Physics Laboratory, Laurel, MD, USA}

\author[0000-0003-3240-7510]{David G. Sibeck}
\affiliation{NASA Goddard Space Flight Center, Greenbelt, MD, USA}

\author[0000-0002-2238-109X]{Laura Vuorinen}
\affiliation{Department of Physics and Astronomy, Queen Mary University of London, Mile End Road, London E1 4NS, UK}

\author[0000-0003-2227-1242]{Adam Michael}
\affiliation{The Johns Hopkins Applied Physics Laboratory, Laurel, MD, USA}

\author[0000-0002-0437-7521]{Mykhaylo Shumko}
\affiliation{The Johns Hopkins Applied Physics Laboratory, Laurel, MD, USA}

\author[0000-0003-2555-5953]{Adnane Osmane}
\affiliation{Department of Physics, University of Helsinki, Helsinki, Finland}

\author[0000-0002-1826-3613]{Eva Krämer}
\affiliation{Department of Physics, Umeå University, 901 87 Umeå, Sweden}

\author[0000-0002-7576-3251]{Lucile Turc}
\affiliation{Department of Physics, University of Helsinki, Helsinki, Finland}

\author[0000-0003-1270-1616]{Tomas Karlsson}
\affiliation{Division of Space and Plasma Physics - KTH Royal Institute of Technology, Stockholm, Sweden}

\author[0000-0002-0604-697X]{Christos Katsavrias}
\affiliation{Department of Physics, National and Kapodistrian University of Athens, Athens, Greece}

\author[0000-0002-4313-1970]{Lynn B. Wilson III}
\affiliation{NASA Goddard Space Flight Center, Greenbelt, MD, USA}

\author[0000-0002-2234-5312]{Hadi Madanian}
\affiliation{NASA Goddard Space Flight Center, Greenbelt, MD, USA}
\affiliation{Catholic University of America, Washington, DC, USA}

\author[0000-0001-7171-0673]{Xóchitl Blanco-Cano}
\affiliation{Departamento de Ciencias Espaciales, Instituto de Geofísica, Universidad Nacional Autónoma de México, Mexico City, Mexico}

\author[0000-0002-9163-6009]{Ian J. Cohen}
\affiliation{The Johns Hopkins Applied Physics Laboratory, Laurel, MD, USA}

\author[0000-0003-4475-6769]{C. Philippe Escoubet}
\affiliation{ESA/ESTEC, Noordwijk, The Netherlands}

\begin{abstract}
Shock-generated transients, such as hot flow anomalies (HFAs), upstream of planetary bow shocks, play a critical role in electron acceleration. Using multi-mission data from NASA's Magnetospheric Multiscale (MMS) and ESA’s Cluster missions, we demonstrate the transmission of HFAs through Earth’s quasi-parallel bow shock, associated with acceleration of electrons up to relativistic energies. Energetic electrons, initially accelerated upstream, are shown to remain broadly confined within the transmitted transient structures downstream, where betatron acceleration further boosts their energy due to elevated compression levels. Additionally, high-speed jets form at the compressive edges of HFAs, exhibiting a significant increase in dynamic pressure and potentially contributing to driving further localized compression. Our findings emphasize the efficiency of quasi-parallel shocks in driving particle acceleration far beyond the immediate shock transition region, expanding the acceleration region to a larger spatial domain. Finally, this study underscores the importance of multi-scale observational approach in understanding the convoluted processes behind collisionless shock physics and their broader implications.
\end{abstract}


\section{Introduction} \label{sec:intro}

Shock-generated transients are kinetic plasma phenomena inherent to collisionless bow shocks, forming through interactions between shock reflected particles and the upstream driving plasma. At planetary quasi-parallel bow shocks (where the angle $\mathrm{\theta_{\mathbf{B}n}}$ between the upstream magnetic field and the shock normal is less than 45$^{\circ}$), various transients can develop under steady-state and varying upstream conditions. This work focuses on the so-called driven shock-generated transients. Driven here refers to the presence of a discontinuity in the magnetic field, while the shock-generated implies the inclusion of both upstream and downstream transients. Having said that, we will first discuss upstream foreshock transients which are typically mentioned in the literature as foreshock/dayside transients \citep{zhang2022dayside} or Transient Upstream Mesoscale Structures (TUMS) \citep{kajdic2024transient}, while typically referring to phenomena such as foreshock bubbles (FBs) \citep{omidi2010foreshock}, hot flow anomalies (HFAs) \citep{schwartz1985active}, and spontaneous hot flow anomalies (SHFAs) \citep{zhang2013spontaneous}. Upstream shock-generated transients, due to their unique properties and size, are very efficient in accelerating electrons to relativistic regime even before they reach Earth's bow shock \citep{wilson2016relativistic,liu2019relativistic,an2020formation,shi2023evidence,raptis2024low}. Another set of transient phenomena that we discuss in this work are high-speed downstream jets corresponding to localized dynamic pressure enhancements in the magnetosheath. Magnetosheath jets are typically formed by kinetic processes at the quasi-parallel shock, \citep{hietala2009supermagnetosonic,plaschke2018jets,suni2021connection,omelchenko20213d,raptis2022downstream,Osmane2024} while they are able to further accelerate electrons within the magnetosheath \citep{liu2020electron,vuorinen2022monte}.

However, what happens when these transients interact with the bow shock is still largely unknown, with only a few observational and simulation studies looking at the interaction of such structures with respect to the shock or its downstream plasma \citep{vsafrankova2002structure,eastwood2008themis,lin2002global,zhao2015case}. More importantly, the effect of their transmission downstream of the shocks with respect to the various processes such as particle acceleration, or generation of high-speed downstream flows is to this date still a mystery.

In this work, we highlight a case study involving two missions in a fortuitous conjunction, NASA's MMS \citep{burch2016magnetospheric} and ESA's Cluster \citep{escoubet2001introduction} observations. During this conjunction, Cluster is located upstream of Earth's bow shock, observing a series of upstream transients. MMS on the other hand is located dawnward from Cluster, observing plasma downstream of the subsolar shock associated to these transients. By analyzing the observations of both missions, we demonstrate that as the upstream transients are transmitted into the magnetosheath, the already energized electrons inside them are additionally accelerated and that the transmission of the compressive edges leads to formation of magnetosheath jets. 

We interchangeably use the terms foreshock transients,  shock-generated transients, and HFAs as the properties demonstrated here are not necessarily an exclusive phenomenon of HFAs but potentially applicable to other upstream mesoscale transients as well.
\section{Data}

The near-Earth observations are obtained from the MMS and Cluster missions. Downstream of the shock, we use data from MMS and employ srvy/fast resolution Level 2 data from the Fast Plasma Investigation \citep[FPI;][]{pollock2016fast} for ion and electron plasma moments, with a temporal resolution of 4.5 seconds. Magnetic field measurements are taken from the Flux Gate Magnetometer \citep[FGM;][]{russell2016magnetospheric}, which provides data every 0.0625 seconds. For energetic electron observations, we use data from the Fly's Eye Energetic Particle Spectrometer \citep[FEEPS;][]{blake2016fly}, taken at 2.5 second cadence. From the Cluster spacecraft, we use magnetic field data from the Fluxgate Magnetometer \citep[FGM;][]{balogh1997cluster}, with a 0.2 second resolution. Proton data are provided from the Composition and Distribution Function (CODIF) spectrometer, a component of the Cluster Ion Spectrometer experiment \citep[CIS;][]{reme1997cluster}, providing measurements approximately every 4 seconds. Electron distributions in the low to suprathermal energy range are obtained from the Plasma Electron and Current Experiment \citep[PEACE;][]{johnstone1997peace} taken approximately every 12.5 seconds. Energetic electron measurements are retrieved from the Research with Adaptive Particle Imaging Detectors \citep[RAPID;][]{wilken2001first}, which provides data approximately every 4 seconds. We make additional use of upstream magnetic field data from the ACE \citep{stone1998advanced}, Wind \citep{wilson2021quarter}, and DSCOVR \citep{burt2012deep} spacecraft, all providing solar wind measurements near the Lagrange 1 (L1) point.  For the MMS data, we focus on MMS1 observations, as its tetrahedral formation, combined with the absence of burst-mode data, results in nearly identical measurements across all MMS spacecraft. Similarly, the Cluster data are exclusively derived from Cluster 4, the only spacecraft that provided proton and electron data at that time.

\section{Results}

\begin{figure*}[ht]
    \centering
{\includegraphics[width=1\textwidth]{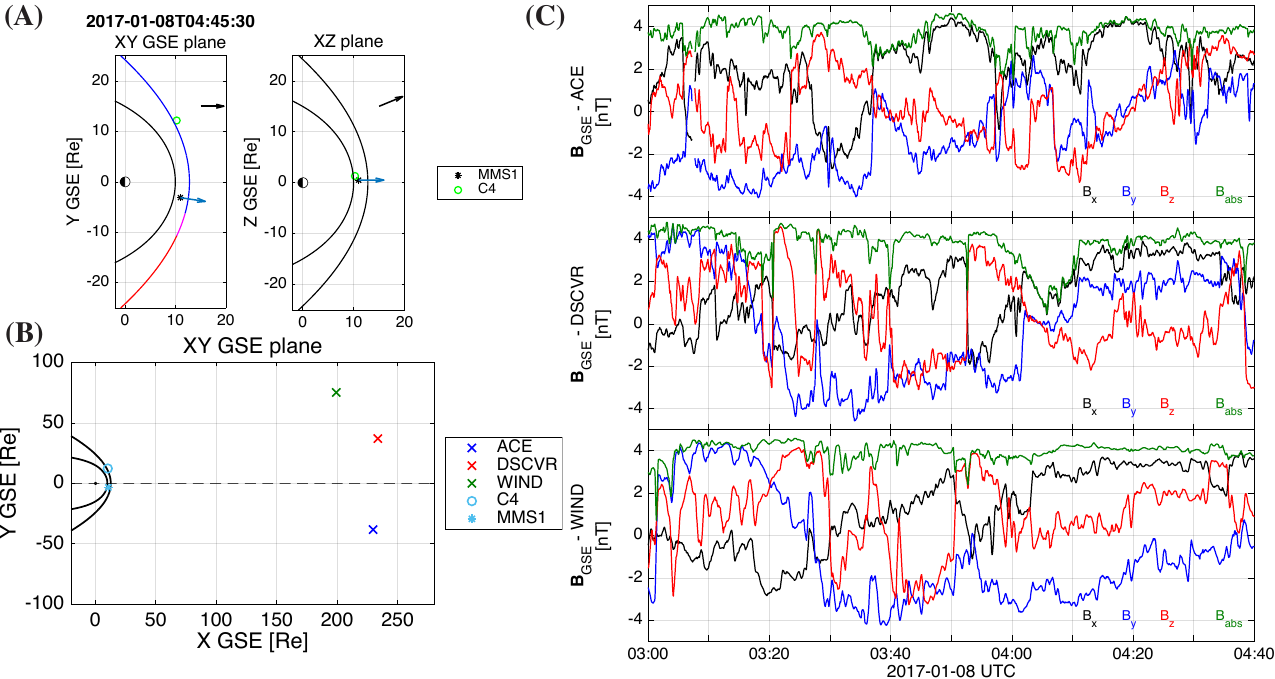}}\caption{
\textbf{(A)} The location of MMS1 and C4 spacecraft in the GSE XY and XZ planes are shown. The bow shock and magnetopause boundary surfaces are shown using \citet{chao2002models} model. The bow shock is color-coded based on 1-hour OMNIweb IMF data, indicating regions of quasi-parallel (blue, $\mathrm{\theta_{\mathbf{B}n}}<40^\circ$), oblique (magenta,  $40^\circ<\mathrm{\theta_{\mathbf{B}n}}<50^\circ$), and quasi-perpendicular (red, $\mathrm{\theta_{\mathbf{B}n}}>50^\circ$) shock. The blue arrow shows the shock normal direction at the closest point of the bow shock model from the MMS location, while the black arrow indicates the direction of the IMF. \textbf{(B)} shows the XY GSE plane along with the positions of the L1 solar wind monitors: ACE, DSCOVR, and WIND. The Sun-Earth line is marked with a dashed line for reference. \textbf{(C)} shows the solar wind magnetic field measurements taken prior to and during the analysis period, highlighting the presence of multiple discontinuities and mesoscale variability even across the $\sim$50 earth radii separation between the L1 solar wind monitors. Such magnetic field configurations are expected to generate disturbances at Earth, contributing to the formation of multiple upstream transient phenomena.}
\label{fig1}
\end{figure*}

Figure \ref{fig1}A and B display the positions of the near-Earth spacecraft (SCs) and the satellites orbiting at L1, while Figure \ref{fig1}C illustrates the interplanetary magnetic field (IMF) components and magnitude measured by the three satellites in the upstream solar wind. The data reveal significant variability in the magnetic field, indicating numerous directional discontinuities that occurred 30-60 minutes before the near-Earth observations discussed below. This mesoscale variability is further evidenced by the different magnetic field variations across the SCs, each of which is located $\gtrsim 10$ Earth radii (Re) apart and offset from the Sun-Earth line. Such variability is characteristic of fast solar wind conditions ($\sim$ 700 km/s), highlighting the complexity and non-planar nature of the solar wind \citep{tsurutani1999review,borovsky2018spatial,borovsky2021solar}. The multiple IMF directional discontinuities are expected to trigger multiple transients on the dayside as soon as they interact with  Earth's bow shock and foreshock \citep{zhang2022dayside,kajdic2024transient}. Based on the relative positions of MMS and Cluster (Figure \ref{fig1}A), we expect Cluster to be located just upstream of the nominal bow shock, while MMS is closer to the subsolar region, downstream of the bow shock.

\begin{figure*}[ht]
    \centering
{\includegraphics[width=0.5\textwidth]{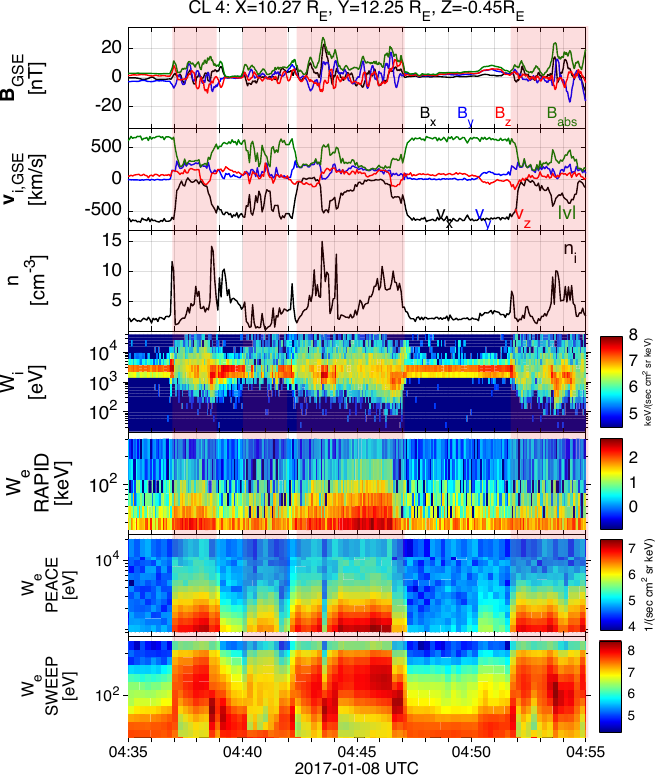}}\caption{
Overview of Cluster observations in the upstream region. From top to bottom, resampled magnetic field in geocentric solar ecliptic (GSE) coordinates (averaged over a 150 data-point window), ion velocity in GSE coordinates, ion density, ion differential energy spectrum, and differential energy spectra for electrons in high-energy, mid-energy, and low-energy ranges. The red-shaded area approximately denotes the core region of multiple foreshock transients that develop upstream of Earth's bow shock.}
\label{fig2}
\end{figure*}

Figure \ref{fig2} provides an overview of upstream Cluster observations closer to Earth, highlighting a series of shock-generated transients in the solar wind. Although some transients observed are challenging to identify and fully characterize, most exhibit typical properties of hot flow anomalies, including deflected and decelerated flows, a heated core, and compressive edges \citep{zhang2022dayside}. The relatively high density within the core region can be attributed to a significant portion of the magnetosheath plasma moving out into the core region \citep{otto2021bow,liu2024observations}. Notably, the fifth panel of Figure \ref{fig2}  shows energetic electrons reaching up to $\sim$200 keV. This relativistic electron population is detected within the foreshock transients under fast solar wind conditions (SW speed is about 700 km/s as shown in the 2nd panel of Figure \ref{fig2})), consistent with recent predictions by \citet{raptis2024low}. Additionally, these energetic electrons appear to be confined within the transient structures, primarily between their compressive edges, with background noise levels dominating the adjacent intervals of solar wind, such as between 04:48 and 04:51. It should be noted that the main focus of this work is not on the precise characterization of the multiple foreshock transient but rather on examining their impact, specifically in terms of their interaction with the shock, their transmission into the magnetosheath, and their role in electron acceleration. Given that the majority of the observed properties align with those of a HFA, we will use this term in the following sections.

\begin{figure*}[ht]
    \centering
{\includegraphics[width=1\textwidth]{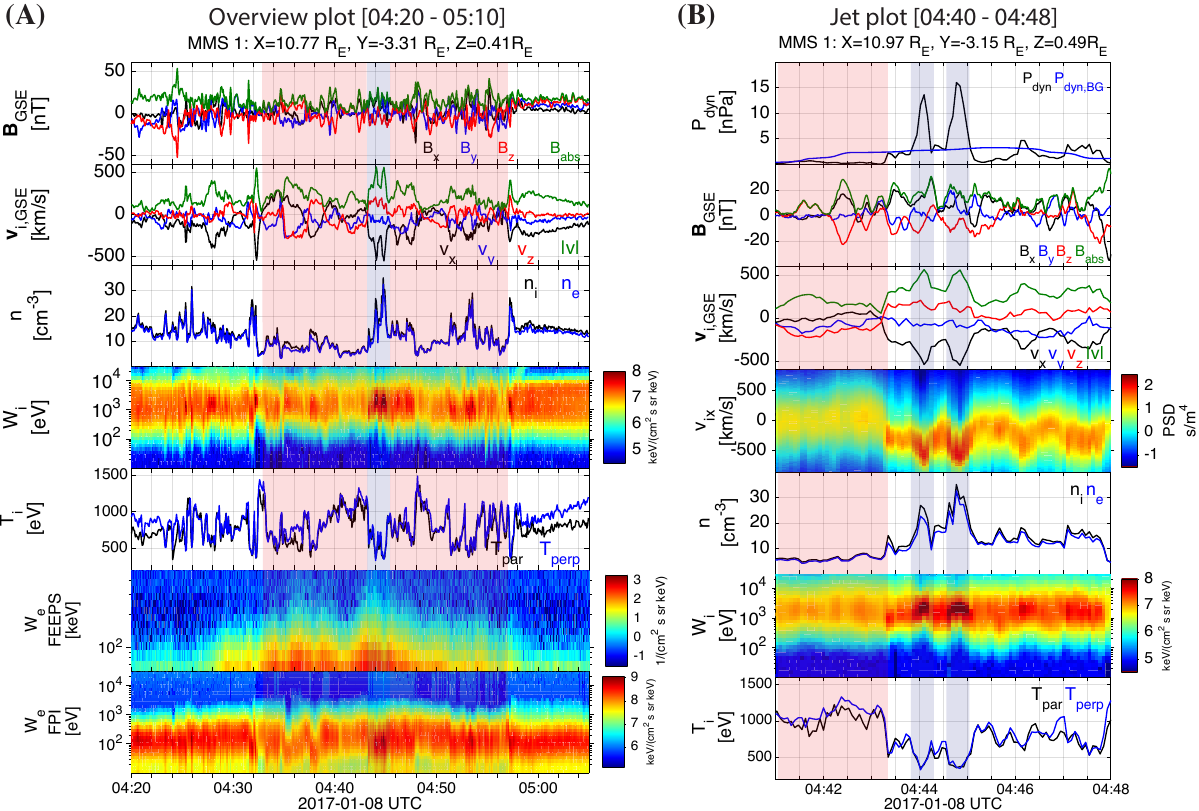}}\caption{
MMS observations of the downstream magnetosheath region. \textbf{(A)} From top to bottom, the panels display the magnetic field resampled in geocentric solar ecliptic (GSE) coordinates and averaged with a 150 data-point window, ion velocity in GSE coordinates, ion and electron densities, the omnidirectional ion energy spectrum from FPI, ion temperatures parallel and perpendicular to the magnetic field, the high-energy electron differential energy spectrum from FEEPS, and the electron differential energy spectrum from FPI. The red-shaded area indicates a thermalized region associated with the ``core'' of the transmitted HFAs, while the blue-shaded area reveals localized high-speed jets forming at the transmitted compressive boundaries. \textbf{(B)} corresponds to a detailed plot of the compressive edge/jet and its adjacent heated core region. From top to bottom, the plot includes ion dynamic pressure, with the background value plotted as a 5-minute window moving average, the magnetic field resampled in GSE coordinates and averaged over a 150 data-point (30 sec) window, ion velocity in GSE coordinates, integrated 1D distribution along the sun-earth line, ion and electron densities, ion differential energy spectrum, and ion temperature. As demonstrated, the localized jets exhibit a relative increase in both density and speed, resulting in dynamic pressure up to 10 times higher than the magnetosheath background.}
\label{fig3}
\end{figure*}

After identifying the presence of relativistic electrons upstream of Earth's bow shock, we now turn our attention to the downstream plasma. Figure \ref{fig3} presents MMS observations from a comparable time frame, accounting for a temporal delay of several minutes. This delay aligns with the substantial spatial separation between the two spacecraft (over 10 $\mathrm{R_{E}}$ along the Y GSE axis), which influences the relative timing of upstream and downstream measurements. Determining a precise one-to-one correspondence between individual upstream transients and their downstream counterparts is beyond the scope of this study. Instead, we approach the whole magnetosheath interval as a disturbed region modulated by the presence of multiple upstream transients. This perspective allows us to focus on their collective properties and their transmission downstream, offering a broader understanding of these phenomena.

As shown in Figure \ref{fig3}, the shock-generated transients within the magnetosheath display typical features of compressed solar wind, such as thermalized ion distributions, magnetic field compression, and flow anomalies. However, key properties of HFAs persist downstream, evidenced by flow irregularities (e.g., strong deviations from earthward flow), suprathermal heated cores, and localized compressions (highlighted in the shaded regions of Figure \ref{fig3}A).

Two key findings emerge from these observations. First, the energetic electrons detected upstream are found at approximately the same interval downstream. This indicates a mechanism that keeps these particles broadly speaking within the localized substructure of the HFAs, preventing them from diffusing into the broader magnetosheath outside the HFAs. While this result provides nearly 30 minutes of continuous observations of high-energy ($>$100 keV) electrons within the magnetosheath, these elevated fluxes are modulated by the intermittent presence of transmitted downstream structures. Thus, they should be interpreted cautiously as they primarily reflect the influence of multiple transient events rather than a single, uniform downstream region of enhanced acceleration. Second, while compressive edges are transmitted downstream as expected, they are also associated with localized increases in Earthward velocity. This observation reveals a colder and faster-moving plasma component (see Fig. \ref{fig3} blue-shaded regions) associated with these transmitted compressive edges, effectively forming high-speed jets downstream of the bow shock shown in Figure \ref{fig3}B.

\begin{figure*}[ht]
    \centering
{\includegraphics[width=1\textwidth]{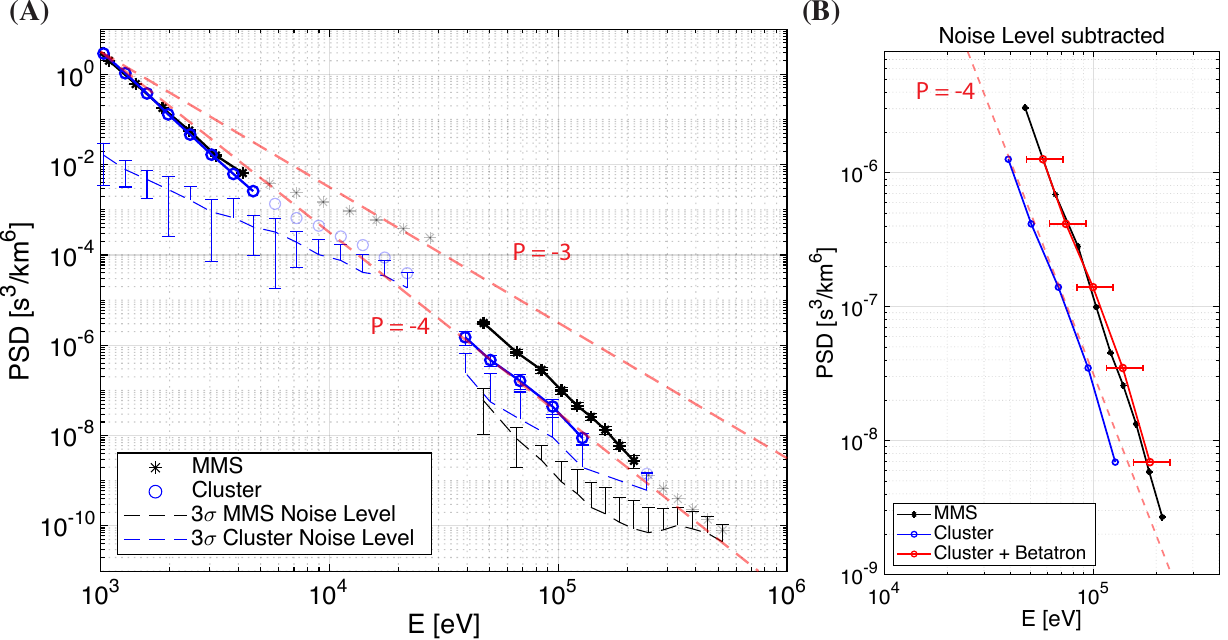}}\caption{
\textbf{(A)} Electron phase space density (PSD) as a function of energy for upstream (Cluster) and downstream (MMS) observations, along with background noise levels for each instrument. Measurements are taken around the energy peak identified by each mission. CL4 data points (blue) are averaged over one minute [04:45–04:46], while MMS measurements (black) are averaged during 30 seconds [04:43:30–04:44:00]. Error bars denote standard deviation (SD), and transparent points indicate values near or below noise levels. Low-energy data are taken from MMS’ FPI and Cluster’s PEACE, while the high-energy tail data is from MMS’ FEEPS and Cluster’s RAPID. The 3$\sigma$ noise levels are shown for both Cluster and MMS high-energy instruments, with Cluster noise taken between [04:47:30–04:49:30] and MMS between [04:00:30–04:09:30]. Red-dashed lines depict spectral slopes ($p$) of -4 and -3. \textbf{(B)} A zoomed-in view of the energetic tail distribution with noise levels subtracted, comparing MMS observations, Cluster observations, and Cluster with betatron acceleration applied. Horizontal error bars show the range of energy gain modeled by betatron acceleration using 1$\sigma$ variability in the upstream and downstream magnetic field measurements taken concurrently with the electron fluxes intervals. The betatron model used is defined as $\mathrm{\Delta E/\langle E_0 \rangle= \frac{2}{3} \Delta B/B_0}$, where $\mathrm{\Delta}$ denotes difference between upstream and downstream measurements, and $\mathrm{B_0,E_0}$ the upstream value.}
\label{fig4}
\end{figure*}

We now examine the relation between the energetic electrons upstream and downstream of the shock. Figure \ref{fig4} compares the electron energy spectra upstream (Cluster) and downstream (MMS), revealing that an additional electron energization occurs during the transmission of HFAs from upstream to downstream of the shock. As shown in Figure \ref{fig4}B, electrons traverse the bow shock experiencing energization consistent with a betatron acceleration under significant scattering ($\mathrm{\Delta B/B_0 \gtrsim 1}$, in which $\mathrm{\Delta B}$ is the difference between the downstream and upstream magnetic field and $\mathrm{B_0}$ is the magnitude of the upstream magnetic field), leading to significant acceleration at higher energies. Upstream magnetic field observations at the compressive edge reach a peak amplitude of about $\sim$30 nT, while downstream measurements show values up to $\sim$45 nT. The betatron model used here originating from the conservation of the first adiabatic moment under efficient pitch angle scattering results in an energy change of $\mathrm{\Delta E/\langle E_0 \rangle= \frac{2}{3} \Delta B/B_0}$, where $\mathrm{\Delta}$ denotes the difference between upstream and downstream measurements and $\mathrm{B_0,E_0}$ the upstream value. (see \cite{liu2019relativistic} for more details on the betatron acceleration model). As depicted in Figure \ref{fig4}B, applying the betatron acceleration model to the energetic electrons observed by Cluster upstream of the shock, results in an almost exact match with the spectrum of the electron population observed by MMS downstream.

This additional compression driving the betatron acceleration can originate from the bow shock transition while potentially being further amplified by localized, time-dependent velocity gradients within the high-speed flows embedded in the compressive edge (see Figure \ref{fig3}B). This is supported by the slight time lag between the density and velocity peaks within the jet interval, a property that has been shown on previous magnetosheath jet observations \citep{raptis2022magnetosheath}. The fact that we observe the density peak(s) occurring before the velocity suggests that the faster flow is situated 'behind' the compression region. This arrangement creates a velocity gradient as the faster-flowing plasma interacts with the denser, slower-moving region ahead of it, potentially driving further localized compression.  Furthermore, the presence of multiple peaks in density can also be associated to the temporal evolution of the compressive edge of the HFA, which acts as a fast shock interacting with its upstream waves \citep{turner2021direct}. Considering that the magnetic field compression ratio from upstream to downstream is approximately $\mathrm{\frac{B_{u}}{B_{d}} \sim 1.8}$ and the density compression ratio is slightly higher at $\mathrm{\frac{n_{u}}{n_{d}} \sim 2.3}$, it is reasonable to assume that the shock transition alone can drive the modeled betatron acceleration. However, as these high-speed flows  approach the magnetopause, their cumulative effect may become more pronounced since there can be more time for localized compression to be driven.

Non-local acceleration mechanisms, such as those linked to particles originating from other regions of the shock or first-order Fermi acceleration between the compressive edge and the magnetopause, may also contribute to the energization of electrons in the magnetosheath \citep{mitchell2014isothermal, schwartz2019collisionless, liu2020electron, vuorinen2022monte}. However, given the concentration of energy flux near the transmitted compressive region (around 04:45) and the presence of already accelerated particles upstream, it is more plausible that the energization occurs primarily at the compressive region itself.

\section{Summary \& Discussion}
Our results can be summarized as follows:
\begin{enumerate} 
\item Multi-spacecraft observations from the MMS and Cluster missions indicate that upstream shock-generated transients, are transmitted through Earth's bow shock, maintaining their characteristic features (deflected and decelerated flows, heated core, localized compressive edges). 
\item Relativistic electrons ( $\sim$200 keV upstream and $\sim$300 keV downstream) are detected in association with foreshock transients, associated with fast solar wind conditions, consistent with the framework proposed by \cite{raptis2024low}.
\item These energetic electrons are modulated by the transmitted transients. Due to the occurrence of multiple transients within a short period, the magnetosheath exhibits nearly continuous elevated fluxes of $>$ 100 keV electrons, observed for up to 30 minutes.
\item The compressive edges of the transmitted foreshock transients exhibit a relative increase in both density and velocity, forming high-speed magnetosheath (downstream) jets characterized by elevated dynamic pressure, similar to the picture shown by \citet{zhou2023undulated,zhou2024magnetosheath}. 
\item The pre-accelerated electrons at upstream transients undergo further energization through betatron acceleration in the downstream region of the bow shock. This compression is primarily driven by the shock transition, while localized velocity gradients associated with high-speed jets may contribute to additional compression downstream.
\end{enumerate}

Overall, our results highlight the critical role that shock-generated transients play in various physical processes.

First, we demonstrate that high-speed downstream jets can form when the compressive edges of HFAs are transmitted downstream of a collisionless shock. Other transients structures (e.g., foreshock cavities and traveling foreshocks) have been associated with the formation of jet-like properties \citep{sibeck2021foreshock,kajdivc2021causes}, while the relation of HFAs to jets been hypothesized and formulated in the past \citep{plaschke2018jets,zhou2023undulated, zhou2024magnetosheath}. However, the one-to-one connection between \textit{in-situ} upstream and downstream observations from a multi-point prospective has not been shown until now. Given that HFAs occur multiple times per day \citep{schwartz2000conditions}, it is possible that a substantial percentage of jets may be linked to HFAs. In particular, the observed jets feature a particularly dense and fast plasma, far exceeding the average properties typically seen in quasi-parallel magnetosheath jets \citep{raptis2020classifying}. This suggests that these jets could have more substantial effects at the magnetopause \citep{ng2021bursty, nvemevcek2023extremely}. Furthermore, our observations are in agreement with previous hybrid simulation results that indicate localized enhancements in density and speed at the compressive edge of shock-generated transients \citep{omidi2016impacts,sibeck2021foreshock}.

Most importantly, we present additional evidence that acceleration from transient phenomena upstream of the shock can generate hundreds of keV electrons at Earth's bow shock environment \citep{raptis2024low}. These electrons, initially energized upstream of the shock, are transmitted downstream and are further accelerated as they cross the bow shock. This additional energization appears to be due to betatron acceleration in which the shocked plasma exhibits compression, which drives particles to even higher energies, while the highly turbulent downstream electromagnetic field allows them to scatter while remaining broadly confined within the transmitted transient region. This supports the understanding that quasi-parallel shocks are efficient particle accelerators, more so than their quasi-perpendicular counterparts \citep{caprioli2014simulations, johlander2021ion, lalti2022database,jebaraj2024direct}. Our findings build upon previous research \citep{wilson2016relativistic, liu2019relativistic, raptis2024low}, which established that foreshock transients alone can accelerate relativistic electrons. We demonstrate that these transients not only drive initial acceleration but also contribute to the sustained energization and confinement of electrons on both sides of the bow shock (upstream and downstream). The almost perfect match between betatron-accelerated upstream electrons and downstream fluxes shows that this process is directly related to particles originating from the upstream region, outside of the planetary magnetosphere system.

Our results also integrate well with previous research on magnetospheric effects. For instance, \citet{katsavrias2021generation} observed downstream jets and Pi2 pulsations\footnote{Pi2 pulsations refer to irregular magnetic field waves occurring within a frequency range of 5-30 mHz. Terminology originates from magnetospheric physics, while Pi2 pulsations have been associated to increased geomagnetic disturbances \citep{shiokawa1998high,keiling2011review}}, with signatures very similar to those reported in this study. Upon revisiting these findings, it becomes clear that these earlier observations were associated with an HFA, which, as in our current work, involved $>$100 keV electrons associated with fast solar wind plasma interacting with Earth's bow shock. In our observations, Pi2 pulsations were detected in the wave-field data of both Cluster and MMS (not shown). This suggests that Pi2 pulsations originating from the HFA core may behave similarly to those reported previously \citep{katsavrias2021generation}, potentially transmitting into the inner magnetosphere and providing evidence for direct foreshock-magnetosphere coupling. 

Overall, the unique multi- mission conjunction event we present in this work emphasizes the importance of adopting a multi-scale approach to address shock-related and particle energization problems underscoring the necessity of obtaining such data. Upcoming missions like HelioSwarm \citep{klein2023helioswarm} will be valuable for this purpose. Additionally, proposed missions such as the Plasma Observatory (PO) mission \citep{retino2021particle}, directly focusing on particle acceleration, and concepts such as Multi-point Assessment of the Kinematics of Shocks (MAKOS) \citep{goodrich2023multi}, which specifically target shocks, will be critical to advance our understanding of collisionless shock physics and associated particle energization.

\section{Conclusion}

In conclusion, this study demonstrates that while quasi-parallel shocks are highly efficient particle accelerators, the primary driver of this acceleration extends beyond the shock transition itself. Rather, it is largely governed by upstream foreshock transients that, through multi-scale processes, energize particles to relativistic levels. Upon transmission into the turbulent downstream region, additional transient phenomena emerge, facilitating betatron acceleration and pushing particles to even higher energy levels. Collectively, these mechanisms expand the effective spatial region for particle acceleration and confinement, making quasi-parallel shocks even more efficient than previously understood.

\begin{acknowledgments}
SR, MS, and IJH were supported by the Magnetospheric Multiscale (MMS) mission of NASA’s Science Directorate Heliophysics Division via subcontract to the Southwest Research Institute (NNG04EB99C). SR acknowledges additional funding support from the Johns Hopkins University Applied Physics Laboratory independent R\&D fund. SR acknowledges the support of the International Space Sciences Institute (ISSI) team 555, “Impact of Upstream Mesoscale Transients on the Near-Earth Environment”, and the Archival Research Visitor Programme of the European Space Agency (ESA). SR acknowledges the support of Zulip (\url{http://zulip.com}) for providing a platform for planning and organization. SR acknowledges the useful discussions with Heli Hietala. ML acknowledges the funding support of the Royal Society awards RF\textbackslash ERE\textbackslash210353, RF\textbackslash ERE\textbackslash 231151, and the STFC grant ST/T00018X/1. DLT is thankful for funding from NASA’s MMS and IMAP missions. DLT also acknowledges funding from National Science Foundation (NSF) Geospace Environment Modeling (GEM) program 2225463. PK's work was funded by the DGAPA PAIIT IN100424 grant. AK acknowledges financial support from NASA's HGIO grant 80NSSC24K0555. EK was supported by Vetenskapsrådet (VR) dnr 2018–03623 and by the Swedish National Space Agency (SNSA) grant 2022-00138. The work of LT is funded by the European Union (ERC grant WAVESTORMS - 101124500). Views and opinions expressed are however those of the author(s) only and do not necessarily reflect those of the European Union or the European Research Council Executive Agency. Neither the European Union nor the granting authority can be held responsible for them. HM was supported by NASA's MMS project, through the Partnership for Heliophysics and Space Environment Research (PHaSER) cooperative agreement. XBC is funded by DGAPA PAPIIT grant IN106724 and CONAHCyt grant  CBF2023-2024-852. AM was supported by the NASA Early Career Investigator Program grant 80NSSC21K0464. \\

We acknowledge the use of cluster observations, accessible through the Cluster Science Archive \citep{laakso2010cluster} \url{https://www.cosmos.esa.int/web/csa)}. Magnetospheric Multiscale (MMS) measurements can be found through \url{https://lasp.colorado.edu/mms/sdc/public/about/browse-wrapper/} or through the Graphical User Interface (GUI) found in \url{https://lasp.colorado.edu/mms/sdc/public/search/}. The location and magnetic field measurements of the Wind, ACE, and DSCOVR mission were taken from Coordinated Data Analysis Web (CDAWeb) \url{https://cdaweb.gsfc.nasa.gov}, while the 1h OMNI solar wind data are accessible via \url{https://omniweb.gsfc.nasa.gov/ow.html}
\end{acknowledgments}

\vspace{5mm}


\software{Python 3.12 with Pyspedas \citep{grimes2022space} was used for downloading and preprocessing data. MATLAB 2024b with IRFU-Matlab \citep{khotyaintsev_2024_11550091} was used for data post-processing, data analysis, and figure generation.}

\bibliography{bib}{}
\bibliographystyle{aasjournal}

\end{document}